\documentclass[prb,twocolumn,showpacs,floatfix,longbibliography]{revtex4-2}
\usepackage{mathrsfs,braket}
\usepackage{amssymb, amsbsy, amsmath, latexsym, dsfont, array, layout, graphicx, mathrsfs, color, ulem, bm}
\usepackage{amsthm}
\usepackage[colorlinks=true,linkcolor=blue,anchorcolor=red,citecolor=blue, urlcolor=blue]{hyperref}

\begin{document}
\title{Probing Floquet topological phases via non-Hermitian skin effect of reflected waves}

\author{Fangqiao Ye\textsuperscript{1, 2}}
\affiliation{\textsuperscript{1}Beijing National Laboratory for Condensed Matter Physics, Institute of Physics, Chinese Academy of Sciences, Beijing 100190, China}
\affiliation{\textsuperscript{2}School of Physical Sciences, University of Chinese Academy of Sciences, Beijing 100049, China}
\author{Haiping Hu\textsuperscript{1, 2}}
\email{hhu@iphy.ac.cn}
\affiliation{\textsuperscript{1}Beijing National Laboratory for Condensed Matter Physics, Institute of Physics, Chinese Academy of Sciences, Beijing 100190, China}
\affiliation{\textsuperscript{2}School of Physical Sciences, University of Chinese Academy of Sciences, Beijing 100049, China}

\begin{abstract}
Periodically driven systems host topological phases without static analogs, such as the anomalous Floquet phase characterized by trivial bulk bands yet robust boundary modes. In this work, we investigate the scattering problem of a Floquet Chern insulator and reveal the non-Hermitian skin effect (NHSE) of reflected waves. Using a discrete-time scattering formalism, we demonstrate how the non-Hermitian winding number of the reflection matrix is linked to the bulk Floquet invariant via boundary resonances. This reflected-wave NHSE relies on which quasienergy gap the incident wave resides in, leading to a gap-dependent Goos-Hänchen (GH) shift. We further show that the momentum-integrated GH shift quantitatively yields the Floquet topological invariant of the corresponding gap. Our work highlights a frequency-dependent  NHSE of reflected waves in driven systems and provides a real-space scattering approach to identify non-equilibrium topology.
\end{abstract}
\maketitle

\section{Introduction}
Subjecting a quantum system to time-periodic driving provides a powerful knob to engineer out-of-equilibrium phases of matter \cite{Bukov2015, Eckardt2017, Oka2019, Oka2009, Lindner2011, Kitagawa2010,Jiang2011,Cayssol2013,Goldman2014,Usaj2014}. This technique, generally known as Floquet engineering, has enabled the realization of topological band structures in various synthetic platforms \cite{Wang2013, McIver2020, Roushan2017, Peng2016,Jotzu2014, Aidelsburger2015, Wintersperger2020,Rechtsman2013,Maczewsky2017,Mukherjee2017}, including ultracold atoms \cite{Jotzu2014, Aidelsburger2015, Wintersperger2020} and photonic crystals \cite{Rechtsman2013,Maczewsky2017,Mukherjee2017}. For a system described by a time-periodic Hamiltonian $H(t)=H(t+\tau)$ with driving period $\tau$, the stroboscopic dynamics are captured by an effective Floquet Hamiltonian $H_F = \frac{i}{\tau}\ln U(\tau)$, where $U(\tau)$ is the time-evolution operator over one period. Because quasienergies are confined to a periodic zone $[-\pi/\tau, \pi/\tau)$, the topological classification of driven systems goes beyond static counterparts \cite{Rudner2013,Titum2016,Nathan2015, Carpentier2015, Roy2017, Yao2017, Morimoto2017,Hu_2020,Huang_2020}. In particular, the dispersion of edge modes may wrap around the quasienergy zone, leading to topological phases with no static analogs \cite{Asboth2014, Fruchart2016}. A primary example is the anomalous Floquet Chern insulator \cite{Rudner2013, Mukherjee2017, Maczewsky2017, Wintersperger2020}, where chiral edge states propagate within the quasienergy gaps even when the Chern numbers of all bulk Floquet bands vanish.

A central issue for periodically driven systems is how to diagnose their topological properties. In experimental platforms such as ultracold atoms, the identification of Floquet topology typically relies on momentum-space state tomography or circular dichroism \cite{Unal2019, Wintersperger2020, Hauke2014, Flaschner2016,Asteria2019}. In practice, reconstructing the bulk invariant involves extensive measurements across the Brillouin zone and precise quantum state manipulation. From a different perspective, the scattering formalism naturally encodes the bulk topology within the scattering matrix \cite{Fulga2012}. In this context, recent studies on static topological insulators have shown that the reflection matrix exhibits the non-Hermitian skin effect (NHSE) \cite{Franca2022}. This effect refers to the directional pumping of eigenstates towards the boundaries, a phenomenon originally studied in non-Hermitian Hamiltonians \cite{Yao2018, MartinezAlvarez2018, Lee2019, Borgnia2020, Okuma2020, Zhang2020, Kunst2018, Yokomizo2019,Hu_2025}. The point-gap topology of the reflection matrix leads to a winding of the reflection phase and induces a transverse Goos-H\"anchen (GH) shift \cite{Franca2022, Ma2020} of the reflected waves. However, these reflection phenomena have only been explored in static systems. Periodically driven systems are distinct: their topology cannot be captured purely by bulk bands, but is instead determined by the full time-evolution operator \cite{Rudner2013, Nathan2015, Carpentier2015, Fruchart2016}. Given this dynamical nature and the cyclic structure of the quasienergy spectrum, the NHSE of reflected waves in Floquet systems remains largely unexplored. A systematic study of such non-equilibrium reflection characteristics offers a potential real-space approach to identify Floquet topology.

In this work, we study the scattering problem of a Floquet Chern insulator and show that its reflection matrix exhibits an energy-dependent NHSE. Within a discrete-time scattering formalism \cite{Fulga2016, Chalker1988, Tajic2005}, we demonstrate that the non-Hermitian winding number of the reflection matrix is directly linked to the bulk Floquet invariant via boundary resonances. Consequently, the reflected wave exhibits the NHSE, and this effect relies on which quasienergy gap the incident wave resides in. Through the stationary phase approximation, we show that the reflection phase winding induces a gap-dependent transverse GH shift on the incident wave. Furthermore, the momentum-integrated GH shift precisely yields the Floquet topological invariant of the corresponding gap. This exact correspondence serves as a reliable signature for Floquet topological phases, including the anomalous phase where all bulk Chern numbers vanish. Our framework provides a real-space probe to identify non-equilibrium topology in driven systems.

The remainder of this paper is organized as follows. Section \ref{secii} introduces a multi-step driving lattice model capable of hosting distinct Floquet topological phases. In Sec. \ref{seciii}, we employ the discrete-time scattering formalism to extract the reflection matrix and establish the equivalence between its non-Hermitian topology and the bulk Floquet invariant via boundary resonances. Section \ref{seciv} presents numerical results for the reflection NHSE and confirms its correspondence with the underlying bulk Floquet topology. In Sec. \ref{secv}, we show that the reflection phase winding induces a gap-dependent GH shift, and the momentum-integrated GH shift recovers the bulk Floquet topological invariant, supported by wave-packet simulations. Finally, Sec. \ref{secvi} summarizes the work.

\section{Floquet Chern insulator model}\label{secii}
\begin{figure}
    \centering
    \includegraphics[width=3.35in]{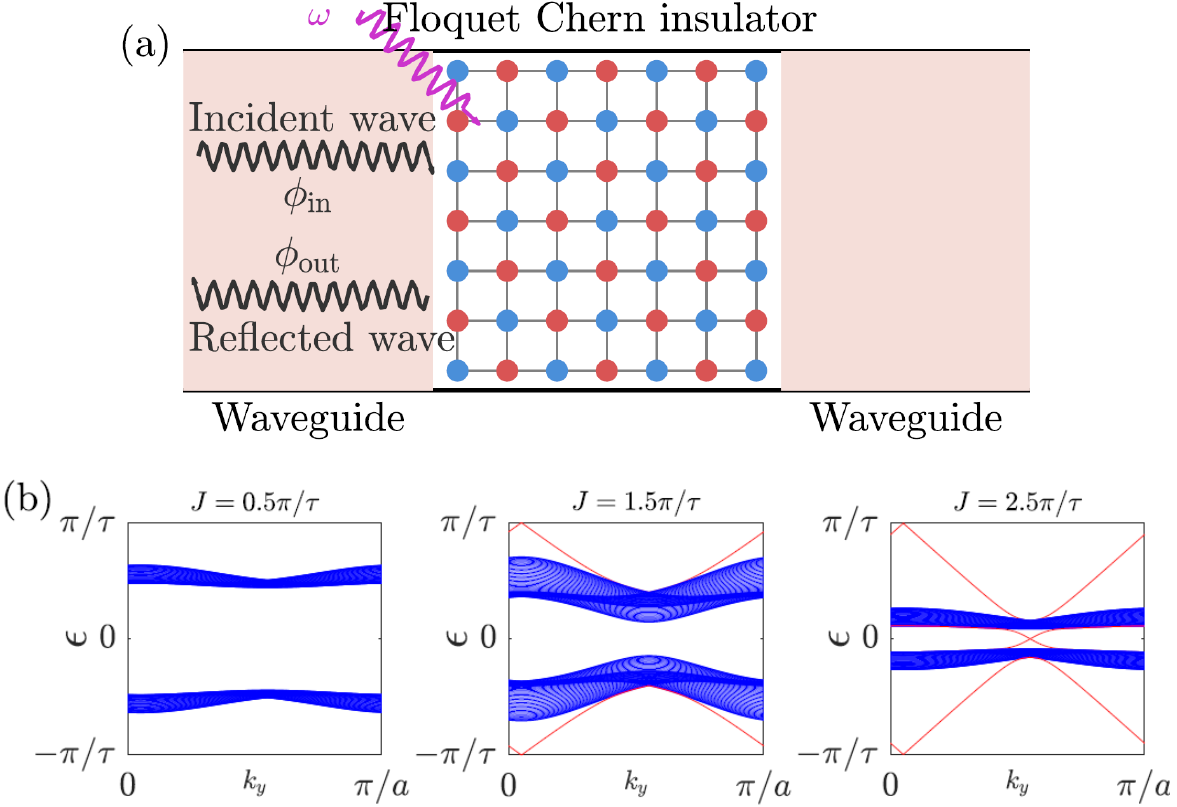}
    \caption{(a) Schematic of the multi-step driving lattice model and scattering settings. (b) Quasienergy band structures $\epsilon(k_y)$ for a strip geometry with open boundary conditions along the $x$ direction and periodic boundary conditions in the $y$ direction. Three representative hopping amplitudes are shown from left to right: $J=0.5\pi/\tau$ [trivial phase with $W_3(0)=0$ and $W_3(\pi)=0$], $J=1.5\pi/\tau$ [Floquet Chern insulator phase, with $W_3(0)=0$ and $W_3(\pi)=1$], and $J=2.5\pi/\tau$ [anomalous Floquet phase, with $W_3(0)=1$ and $W_3(\pi)=1$]. The system size along $x$ direction is $L_x=60$.}
    \label{fig1}
\end{figure}

We consider a prototypical multi-step driving lattice model introduced by Rudner \textit{et al.} \cite{Rudner2013}. As sketched in Fig.~\ref{fig1}(a), the model is defined on a two-dimensional bipartite square lattice consisting of sublattices $A$ and $B$. The system is described by a time-periodic Hamiltonian $H(t) = H(t+\tau)$ with driving period $\tau=1$. The driving protocol consists of five piecewise-constant segments of equal duration $\tau/5$. During the first four segments ($n=1,2,3,4$), the inter-sublattice hopping is turned on exclusively along the $\boldsymbol{b}_n$ direction with an amplitude $J_{n}(t) = J$, while hoppings in all other directions are zero. The nearest-neighbor vectors are $\boldsymbol{b}_{1}=-\boldsymbol{b}_{3}=(a,0)$ and $\boldsymbol{b}_{2}=-\boldsymbol{b}_{4}=(0,a)$, with $a$ being the lattice constant. In the final segment ($n=5$), all hopping terms are switched off ($J_n = 0$), and only a sublattice potential difference $\delta_{AB}(t) = \Delta$ is applied. In momentum space, the Hamiltonian reads:
\begin{equation}
    H(\boldsymbol{k},t) = -\sum_{n=1}^{4} J_{n}(t) \left( e^{i\boldsymbol{b}_{n}\cdot\boldsymbol{k}}\sigma^{+} + e^{-i\boldsymbol{b}_{n}\cdot\boldsymbol{k}}\sigma^{-} \right) + \delta_{AB}(t)\sigma_{z},
\end{equation}
where $\boldsymbol{k}=(k_x, k_y)$ is the crystal momentum, while $\sigma^{\pm}=(\sigma_{x}\pm{i}\sigma_{y})/2$ and $\sigma_z$ are the Pauli matrices acting on the sublattice degree of freedom.

The dynamics of the system is described by the time-evolution operator
\begin{equation}
    U(\boldsymbol{k}, t) = \mathcal{T} \exp \left[ -i \int_0^t H(\boldsymbol{k}, t') dt' \right],
\end{equation}
where $\mathcal{T}$ denotes time ordering. According to the Floquet theorem, for a time-periodic Hamiltonian, the time-evolution operator can be written as $U(\boldsymbol{k}, t) = P_\epsilon(\boldsymbol{k}, t) e^{-i H_F^\epsilon(\boldsymbol{k}) t}$. Here, $H_F^\epsilon(\boldsymbol{k}) = \frac{i}{\tau} \log U(\boldsymbol{k}, \tau)$ is the time-independent effective Floquet Hamiltonian. Specifying this logarithm requires choosing a branch cut at a reference energy $\epsilon$ typically within a bulk gap. Correspondingly, $P_\epsilon(\boldsymbol{k}, t) = P_\epsilon(\boldsymbol{k}, t+\tau)$ is the periodic micromotion operator associated with this choice of $\epsilon$. From a stroboscopic viewpoint, the evolution after one complete driving cycle is given by $U(\boldsymbol{k}, \tau) = e^{-i H_F^\epsilon(\boldsymbol{k}) \tau}$. The eigenvalues of $U(\boldsymbol{k}, \tau)$ define the quasienergy bands $\epsilon(\boldsymbol{k})$, and the corresponding eigenstates are the Floquet modes, satisfying $U(\boldsymbol{k}, \tau)|\psi_{\boldsymbol{k}}\rangle = e^{-i\epsilon(\boldsymbol{k}) \tau}|\psi_{\boldsymbol{k}}\rangle$. Due to the discrete time-translation symmetry, the quasienergies are periodic and restricted to the Floquet Brillouin zone $[-\pi/\tau, \pi/\tau)$.

The cyclic nature of the quasienergy spectrum leads to a breakdown of the standard bulk-edge correspondence available in static systems. In static two-band models, the number of chiral edge modes is given by the bulk Chern numbers. However, in Floquet systems, chiral edge modes can cross the quasienergy gaps at $\epsilon = 0$ or $\epsilon = \pi/\tau$ even if the Chern numbers of all bulk bands are zero. In fact, Floquet topological phases are non-equilibrium phases. To characterize their topological properties, one should utilize the winding number $W_3(\epsilon)$ \cite{Rudner2013}, defined in the (2+1)-dimensional momentum-time space as:
\begin{equation}
\begin{aligned}
    W_3(\epsilon) &= \frac{1}{8\pi^{2}}\int dt dk_x dk_y \\
    &\quad \times \text{Tr} \left( P_{\epsilon}^{-1}\partial_t P_{\epsilon} \left[ P_{\epsilon}^{-1}\partial_{k_x}P_{\epsilon}, P_{\epsilon}^{-1}\partial_{k_y}P_{\epsilon} \right] \right).
\end{aligned}
\end{equation}
The number of chiral edge modes residing in a specific gap $\epsilon$ is given by $n_{\text{edge}}(\epsilon) = W_3(\epsilon)$.

In Fig.~\ref{fig1}(b), we show the quasienergy band structures of three distinct Floquet topological phases of the model by tuning the hopping amplitude $J$. We adopt a strip geometry with periodic boundary condition (PBC) along the $y$ direction and open boundary condition (OBC) along the $x$ direction. (i) When $J=0.5\pi/\tau$, the system is in a topologically trivial phase. Both the $0$-gap and the $\pi$-gap have zero winding numbers ($W_3(0)=0, W_3(\pi)=0$), resulting in the complete absence of chiral edge modes. The Chern numbers of the lower and upper bands are $C_-=C_+=0$. (ii) When $J=1.5\pi/\tau$, the system enters a Floquet Chern insulator phase. The winding numbers are $W_3(0)=0$ and $W_3(\pi)=1$, and chiral edge modes appear exclusively in the $\pi$-gap. The Chern numbers of the two bulk bands are $C_-=-1$ and $C_+=1$. (iii) When $J=2.5\pi/\tau$, the system transitions into the anomalous topological phase. In this regime, there exist chiral edge states in both gaps, captured by the winding numbers $W_3(0)=1$ and $W_3(\pi)=1$. Note that the Chern numbers of both bulk bands vanish ($C_-=C_+=0$).

\section{Discrete-time scattering formalism and reflection topology}\label{seciii}

In this section, we employ a discrete-time scattering formalism to probe the topology of the driven lattice. The explicit scattering matrix is first derived for a wave packet incident on the boundary of the two-dimensional lattice. Subsequently, an analysis of the reflection phase winding reveals its exact correspondence with the underlying bulk Floquet phases.

\subsection{Scattering matrix}
We consider a scattering setup by coupling the periodically driven lattice to two semi-infinite, topologically trivial leads at its left ($x=1$) and right ($x=L_x$) boundaries, as sketched in Fig.~\ref{fig1}(a). The scattering region is periodic in the $y$ direction with $L_y$ lattice sites and finite in the $x$ direction with $L_x$ lattice sites. We employ the discrete-time scattering formalism \cite{Fulga2016, Chalker1988}, which provides a stroboscopic description of the scattering process at an incident energy $\epsilon$. Under the wide-band limit, the leads are treated as memoryless reservoirs with an energy-independent density of states. Consequently, the coupling between the leads and the system becomes instantaneous, effectively occurring at integer multiples of the driving period $\tau$. Between these discrete updates, the internal state undergoes a full period of evolution governed by the bulk Floquet operator $U$.

Let $\phi^{\text{in}} = (\phi^{\text{in}}_L, \phi^{\text{in}}_R)^T$ and $\phi^{\text{out}} = (\phi^{\text{out}}_L, \phi^{\text{out}}_R)^T$ denote the incoming and outgoing wave amplitudes in the left ($L$) and right ($R$) external leads, respectively. We introduce a projection operator $\Pi$ that maps the internal lattice degrees of freedom onto the boundary sites coupled to the leads. The operator $\Pi$ is a $2L_y \times L_x L_y$ matrix. Its matrix elements are explicitly given by $\Pi_{m, (x,y)} = \delta_{m, y}\delta_{x, 1} + \delta_{m, L_y+y}\delta_{x, L_x}$, where the channel index $m \in \{1, \dots, 2L_y\}$ ($1 \le m \le L_y$ labels the left lead, and $L_y+1 \le m \le 2L_y$ labels the right lead), and $(x,y)$ are the spatial coordinates of the lattice sites with $x \in \{1, \dots, L_x\}$ and $y \in \{1, \dots, L_y\}$. To determine the scattering matrix $S(\epsilon)$ defined by $\phi^{\text{out}} = S(\epsilon) \phi^{\text{in}}$, we relate the external lead amplitudes to the internal wave function and derive the self-consistent condition induced by the stroboscopic dynamics. Let $\psi$ be the internal wave function at the end of a driving cycle. At this instant, the components $\Pi \psi$ escape into the leads to form the outgoing wave $\phi^{\text{out}} = \Pi \psi$, while the remaining part $(1 - \Pi^T \Pi) \psi$ stays in the system. Simultaneously, the input $\phi^{\text{in}}$ is injected from the leads. Thus, the newly formed internal state $\tilde{\psi}$ at the beginning of the next driving cycle is:
\begin{equation}
    \tilde{\psi} = (1 - \Pi^T \Pi) \psi + \Pi^T \phi^{\text{in}}.
\end{equation}
This state then evolves for one full driving period governed by the Floquet operator $U$.

For a steady scattering process at an incident energy $\epsilon$, the evolved state $U \tilde{\psi}$ at the end of this new cycle should be identical to the previous end-of-cycle state $\psi$, up to a dynamical phase factor $e^{-i\epsilon \tau}$. This gives a self-consistent equation relating the internal state to the incident wave:
\begin{equation}
    U [(1 - \Pi^T \Pi) \psi + \Pi^T \phi^{\text{in}}] = e^{-i\epsilon \tau} \psi.
\end{equation}
Solving this equation for the internal wave function $\psi$, we have $\psi = [e^{-i\epsilon \tau} - U(1 - \Pi^T \Pi)]^{-1} U \Pi^T \phi^{\text{in}}$. The outgoing wave amplitudes $\phi^{\text{out}}$ are extracted from this end-of-cycle state via $\phi^{\text{out}} = \Pi \psi$. Substituting the expression for $\psi$ into this relation yields:
\begin{equation}
    \phi^{\text{out}} = \Pi [e^{-i\epsilon \tau} - U(1 - \Pi^T \Pi)]^{-1} U \Pi^T \phi^{\text{in}}.
\end{equation}
By factoring out $e^{-i\epsilon \tau}$ from the inverse matrix, we obtain the explicit expression for the energy-dependent scattering matrix $S(\epsilon)$:
\begin{equation}
    S(\epsilon) = \Pi [1 - e^{i\epsilon \tau} U (1 - \Pi^T \Pi)]^{-1} e^{i\epsilon \tau} U \Pi^T. \label{eq:s_matrix}
\end{equation}
Equation~\eqref{eq:s_matrix} admits an intuitive physical interpretation by expanding the inverse matrix as a geometric series: $[1 - e^{i\epsilon \tau} U Q]^{-1} = \sum_{n=0}^{\infty} [e^{i\epsilon \tau} U Q]^n$, where $Q \equiv 1 - \Pi^T \Pi$ acts as a projector for internal reflections. In this multiple-scattering picture, the total scattering matrix is a coherent superposition of processes involving different dwelling times. The zeroth-order term ($n=0$) represents prompt scattering: the wave is injected by $\Pi^T$, evolves for a single driving period via $U$ while accumulating a dynamical phase $e^{i\epsilon \tau}$, and immediately escapes through $\Pi$. Higher-order terms ($n \ge 1$) describe delayed scattering events where the wave undergoes $n$ internal bounces off the boundaries (enforced by $Q$) and spends $n+1$ periods exploring the bulk before finally escaping.

\subsection{Topology of reflection waves}

The scattering matrix $S(\epsilon)$ is a $2L_y \times 2L_y$ matrix that can be partitioned into reflection and transmission blocks:
\begin{equation}
    S(\epsilon) = \begin{pmatrix} S_{ll}(\epsilon) & S_{lr}(\epsilon) \\ S_{rl}(\epsilon) & S_{rr}(\epsilon) \end{pmatrix}.
\end{equation}
We focus on the upper-left $L_y \times L_y$ block $S_{ll}(\epsilon)$ (hereafter denoted as $R$), which describes the reflection of left-incident waves back into the left lead. With PBC applied in the $y$ direction, the transverse momentum $k_y$ is conserved across the interface, and the reflection matrix is characterized by $k_y$ as $R(\epsilon, k_y)$. When the incident energy $\epsilon$ lies in a Floquet gap, the absence of propagating bulk modes causes transmission to be exponentially suppressed in a sufficiently long system ($L_x \to \infty$). In this regime, the incident waves are totally reflected, and $R(\epsilon, k_y)$ becomes a unitary matrix. Its eigenvalues can thus be written as $e^{i\theta_\alpha(\epsilon, k_y)}$, where $\theta_\alpha(\epsilon, k_y)$ are the reflection phases. As $k_y$ sweeps across the 1D transverse Brillouin zone, these eigenvalues trace out closed trajectories in the complex plane. The phase winding of the reflection matrix is characterized by a one-dimensional winding number:
\begin{equation}
    \nu(\epsilon) = \frac{1}{2\pi i} \int dk_y \partial_{k_y} \ln \det [R(\epsilon, k_y)].
\end{equation}

We now relate this invariant to the bulk Floquet topology. To this end, we consider a semi-infinite geometry extending in the $x \ge 1$ direction to focus on the scattering dynamics at the left boundary. This setup isolates the left boundary resonances and eliminates finite-size interference from the right boundary. Let $\Pi_l$ be the projection operator restricted to the left boundary sites, and $Q_l = 1 - \Pi_l^T \Pi_l$ be the projector onto the remaining internal sites. Using these operators, the reflection matrix $R(\epsilon, k_y)$ evaluated at an incident energy $\epsilon$ can be expressed as:
\begin{equation}
    R(\epsilon, k_y) = \Pi_l \left[ 1 - e^{i\epsilon \tau} U(k_y) Q_l \right]^{-1} e^{i\epsilon \tau} U(k_y) \Pi_l^T,
\end{equation}
where $U(k_y)$ is the unitary Floquet operator of the semi-infinite system. The properties of $R(\epsilon, k_y)$ are characterized by its poles in the complex plane. These poles occur at complex energy $z$ where the inverse matrix becomes singular, i.e., $\det\left[ 1 - e^{i z \tau} U(k_y) Q_l \right] = 0$. We define the effective evolution operator for the internal lattice as $U_{\text{eff}}(k_y) = Q_l U(k_y) Q_l$. Finding these poles is equivalent to solving the eigenvalue problem $U_{\text{eff}}(k_y) |\psi\rangle = \lambda |\psi\rangle$, with $\lambda = e^{-i z \tau}$. Because $Q_l$ projects out the left boundary, wave packets inside the lattice lose amplitude as they hit the boundary and escape into the lead. Consequently, $U_{\text{eff}}$ is non-unitary, yielding eigenvalues that satisfy $|\lambda| < 1$. We can thus parameterize the complex pole as $z = \epsilon_r - i\gamma$, where $\epsilon_r$ is the resonance quasienergy and $\gamma > 0$ is the decay width.

A system characterized by a bulk Floquet winding number $W_3(\epsilon)$ supports the same number of chiral edge states within the corresponding quasienergy gap. Upon coupling the left boundary to the lead, these edge states transform into the boundary resonances of $U_{\text{eff}}(k_y)$. As $k_y$ sweeps, these states traverse the gap, creating a net number of $W_3(\epsilon)$ resonant momenta $k_y^*$ where the real part of the resonance quasienergy matches the incident energy: $\epsilon_r(k_y^*) = \epsilon$. According to the argument principle of complex analysis, as $k_y$ passes through each resonant momentum $k_y^*$, the presence of this complex pole causes the total reflection phase $-i\ln \det R(\epsilon, k_y)$ to accumulate a $2\pi$ shift. The reflection winding number $\nu(\epsilon)$ counts the net number of these $2\pi$ phase wraps. This directly corresponds to the number of chiral edge states determined by the bulk invariant $W_3(\epsilon)$ and we obtain the equivalence:
\begin{equation}
    \nu(\epsilon) = W_3(\epsilon).
\end{equation}
Thus, an exact correspondence is established between the bulk Floquet topology and the reflection phase winding via boundary resonances.
\begin{figure}
    \centering
    \includegraphics[width=1\linewidth]{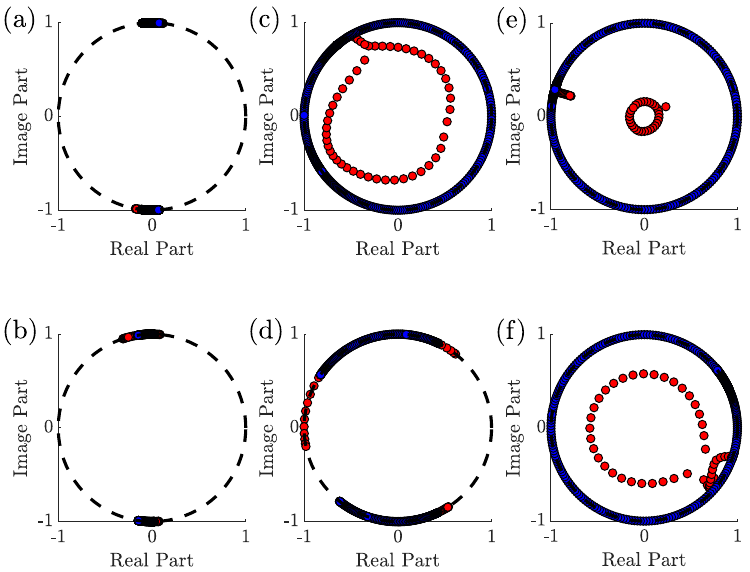}
    \caption{Eigenspectra of the reflection matrix across different Floquet topological phases in the complex plane. Blue and red dots represent the reflection spectra obtained under transverse periodic (strip geometry) and open boundary conditions, respectively. (a),(c),(e) Reflection spectra evaluated for an incident energy $\epsilon=\pi$ for the trivial phase ($J=0.5\pi/\tau$), the Floquet Chern insulator phase ($J=1.5\pi/\tau$), and the anomalous Floquet phase ($J=2.5\pi/\tau$), respectively. (b),(d),(f) Corresponding spectra for the same three phases, but with the incident energy $\epsilon=0$. The system size is $L_x=60$ along $x$ direction.}
    \label{fig2}
\end{figure}

\section{Non-Hermitian skin effect of reflected waves}\label{seciv}

The reflection matrix encoding the bulk Floquet topology is inherently non-Hermitian. We now study its spectral properties under different boundary conditions. We first consider a strip geometry with PBC in the $y$ direction, which conserves the transverse momentum $k_y$. Setting the system length to $L_x=60$, the $k_y$-dependent reflection matrix $R(\epsilon, k_y)$ is a $2\times 2$ matrix due to the bipartite unit cell. We numerically calculate the eigenvalues of $R(\epsilon, k_y)$ for incident energies $\epsilon = 0$ and $\pi$. As required by total reflection within the bulk gap, all eigenvalues lie on the unit circle in the complex plane. However, their phase distributions depend on the underlying Floquet phase, as shown in Fig.~\ref{fig2}. For the trivial phase ($J=0.5\pi/\tau$), the eigenvalues form disconnected arcs for both $\epsilon = 0$ and $\pi$, giving a winding number $\nu(0)=\nu(\pi) = 0$. For $J=1.5\pi/\tau$, the eigenvalues wrap the entire unit circle for $\epsilon$ in the $\pi$-gap ($\nu(\pi)=1$). They only form partial arcs in the $0$-gap ($\nu(0)=0$). For the anomalous Floquet phase ($J=2.5\pi/\tau$), the eigenvalues wrap the full unit circle for both gaps ($\nu(0)=\nu(\pi)=1$). These numerical observations confirm the theoretical correspondence established in Sec.~\ref{seciii}.
\begin{figure}
    \centering
    \includegraphics[width=1\linewidth]{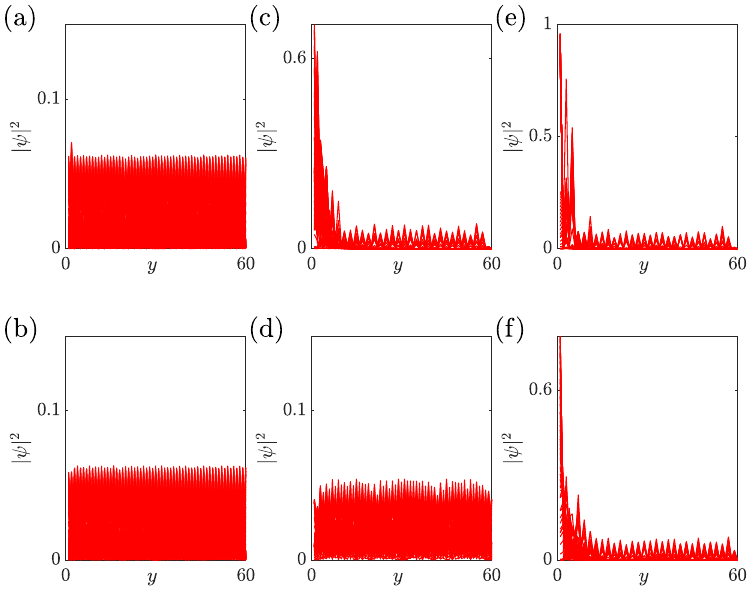}
    \caption{Spatial profiles of the reflection eigenvectors under open boundary conditions. (a),(c),(e) Distributions of all eigenvectors of the reflection matrix for the trivial phase, the Floquet Chern insulator phase, and the anomalous Floquet phase, respectively, evaluated for an incident energy $\epsilon=\pi$. (b),(d),(f) Corresponding profiles for the same three phases, but with the incident energy $\epsilon=0$. The system size is $L_x=L_y=60$.}
     \label{fig3}
\end{figure}

In non-Hermitian physics, a non-zero spectral winding number under PBC signifies a nontrivial point-gap topology \cite{Gong2018,Kawabata2019,Okuma2020} and leads to the NHSE under OBC. To verify this, let us apply OBC in the $y$ direction with lattice sizes $L_x=L_y=60$. The corresponding eigenspectra and eigenvector distributions of $R(\epsilon)$ are shown in Fig.~\ref{fig2} and Fig.~\ref{fig3}. For $J=0.5\pi/\tau$, the reflection spectrum remains on the unit circle for both incident energies $0$ and $\pi$, and the eigenvectors are spatially extended. For $J=1.5\pi/\tau$ with $\epsilon$ in the $\pi$-gap, the spectrum collapses inside the unit circle. The corresponding eigenvectors exhibit the NHSE and localize exponentially at the transverse boundaries. In contrast, when the incident energy is in the $0$-gap ($\nu(0)=0$), the spectrum remains on the unit circle with extended eigenvectors. For the anomalous phase ($J=2.5\pi/\tau$), the spectrum collapses and the NHSE appears for incident energies in both the $0-$ and $\pi-$gaps. 

The collapse of the reflection spectrum and the associated NHSE can be understood through the boundary scattering dynamics. An eigenvalue $|\lambda| < 1$ of the reflection matrix implies a loss of probability amplitude in the reflection process. Under $y$-OBC, the transverse truncation creates corners at the lead-system interface. A system with a nonvanishing invariant $W_3(\epsilon)$ hosts chiral edge states within the corresponding $\epsilon$-gap. When an incident wave couples into these edge channels, it is transported transversely along the $y$-boundaries instead of reflecting immediately back into the lead. This transverse transport reduces the reflection amplitude at the interface and pulls the eigenvalues inside the unit circle. The skin modes are eigenstates of $R(\epsilon)$, whose transverse spatial profiles remain invariant upon reflection. The unidirectional transport by the chiral edge states causes these invariant profiles to localize exponentially at a specific corner. As we will discuss in the next section, this transverse transport along the interface translates into a GH shift for a finite-width wave packet.

\section{Gap-dependent Goos-H\"anchen shift}\label{secv}

The non-Hermitian topology of the reflection matrix provides a real-space scheme to quantitatively determine the invariant $W_3(\epsilon)$. Let us consider a Gaussian wave packet propagating toward the interface, with transverse profile:
\begin{equation}
    \Psi_{\text{in}}(y,t) = \int dk_y f(k_y - k_{y0}) e^{i (k_y y - \epsilon_0 t)},
\end{equation}
where $f(k_y - k_{y0}) = (2\pi \sigma^2)^{-1/4} \exp[-(k_y - k_{y0})^2 / 4\sigma^2]$ is a momentum envelope, $k_{y0}$ is the central momentum, and $\epsilon_0$ is the incident energy. Upon reflection, the wave acquires the momentum-dependent reflection phase $\theta(\epsilon_0, k_y)$. The transverse profile of the reflected wave at the boundary is:
\begin{equation}
    \Psi_{\text{ref}}(y,t) = \int dk_y f(k_y - k_{y0}) e^{i [k_y y - \epsilon_0 t + \theta(\epsilon_0, k_y)]}.
\end{equation}
Assuming the envelope $f$ is narrow, we apply the stationary phase approximation \cite{Beenakker2009, Bliokh2013} and expand the total phase $\Phi = k_y y - \epsilon_0 t + \theta(\epsilon_0, k_y)$ to first order around $k_{y0}$. Due to constructive interference, the spatial center of the wave packet is determined by the condition $\partial \Phi / \partial k_y |_{k_{y0}} = 0$. This yields the spatial center $y_c$ of the reflected wave packet. The resulting transverse spatial translation is the GH shift:
\begin{equation}
    \Delta_{GH}(\epsilon_0, k_{y0}) = y_c = -\frac{\partial \theta(\epsilon_0, k_y)}{\partial k_y}\Bigg|_{k_{y0}}.
\end{equation}
Physically, this spatial shift arises from the same scattering dynamics that lead to the NHSE discussed in Sec.~\ref{seciv}. When the incident wave strikes the interface, it couples to the chiral edge states. Because these states propagate unidirectionally, they transport the wave amplitude transversely along the boundary before reflection.

By integrating this spatial shift over the incident momenta, we obtain the winding number of the reflection matrix:
\begin{equation}
    \int \Delta_{GH}(\epsilon_0, k_{y0}) dk_{y0} = -\int \frac{\partial \theta}{\partial k_y} dk_y = -2\pi \nu(\epsilon_0).
\end{equation}
While the local shift $\Delta_{GH}$ at a specific momentum $k_{y0}$ depends on scattering details, its integration gives the winding number of the reflection matrix. Given the equivalence $\nu(\epsilon_0) = W_3(\epsilon_0)$, the momentum-integrated spatial shift serves as a direct probe of the bulk Floquet topology. This differs from extracting Floquet topological invariants via momentum-space tomography \cite{Hauke2014, Flaschner2016} or measurements based on shaking protocols and circular dichroism \cite{Unal2019, Wintersperger2020,Asteria2019}, which require mapping the bulk states across the entire Brillouin zone. In contrast, our scattering approach confines the measurement to the physical boundary. One prepares incident wave packets with varying initial momenta $k_{y0}$ and records the GH shift of the reflected waves. Distinct quasienergy gaps can be independently probed by simply tuning the incident energy $\epsilon_0$ of the wave packet.

\begin{figure}
    \centering
    \includegraphics[width=1\linewidth]{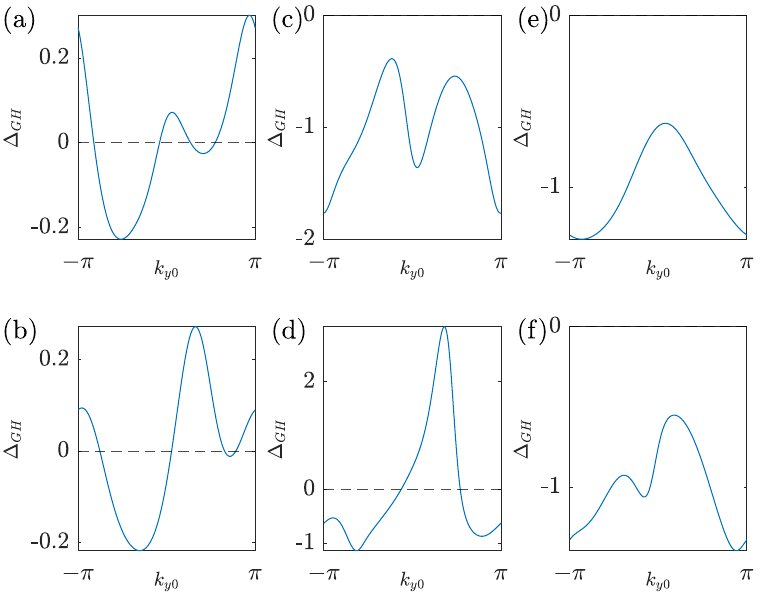}
    \caption{Goos-H\"anchen shift of the reflection waves. Momentum-dependent GH shift $\Delta_{GH}(\epsilon_0, k_{y0})$ as a function of the central transverse momentum $k_{y0}$. (a),(c),(e) GH shifts evaluated for an incident energy $\epsilon_0=\pi$ for the trivial phase ($J=0.5\pi/\tau$), the Floquet Chern insulator phase ($J=1.5\pi/\tau$), and the anomalous Floquet phase ($J=2.5\pi/\tau$), respectively. (b),(d),(f) Corresponding GH shifts for the same three phases, but with the incident energy $\epsilon_0=0$. The numerical simulations are performed with a system size $L_x=L_y=60$ and a wave-packet width $\sigma=6$.}
    \label{fig4}
\end{figure}

We present the numerical evaluation of the GH shift $\Delta_{GH}(\epsilon_0, k_{y0})$ as a function of the transverse momentum $k_{y0}$ for the three representative phases, as shown in Fig.~\ref{fig4}. To simulate the scattering process on a finite lattice, we initialize a real-space Gaussian wave packet $\psi_{\text{in}}(y) = \mathcal{N} \exp[-(y-y_0)^2 / 2\sigma^2] \exp(i k_{y0} y)$, where $\mathcal{N}$ is the normalization constant and $y_0$ is the initial center. By Fourier transform, this real-space initialization corresponds to a momentum envelope centered at $k_{y0}$. Upon applying the reflection matrix $R(\epsilon_0)$, the spatial center of the wave function is given by:
\begin{equation}
    \bar{r} = \sum_{y=1}^{L_y} y \, |\psi(y)|^2.
\end{equation}
The numerical GH shift is then calculated as $\Delta_{GH}(\epsilon_0, k_{y0}) = \bar{r}_{\text{ref}} - \bar{r}_{\text{in}}$, where $\bar{r}_{\text{in}}$ and $\bar{r}_{\text{ref}}$ are the centers of the incident and reflected wave packets, respectively. As shown in Fig.~\ref{fig4}, for the trivial phase ($J=0.5\pi/\tau$), the GH shift varies with $k_{y0}$, but its integral over all initial momenta vanishes for incident energies in both the $0$ and $\pi$ gaps. For $J=1.5\pi/\tau$, the numerical integration of $\Delta_{GH}(\epsilon_0, k_{y0})$ yields $-2\pi$ when the incident energy resides in the $\pi$-gap, corresponding to $W_3(\pi) = 1$, whereas it vanishes for the $0$-gap. For the anomalous Floquet phase ($J=2.5\pi/\tau$), the integration of the GH shift yields $-2\pi$ for incident energies in both the $0$ and $\pi$ gaps ($W_3(0)=W_3(\pi)=1$). These numerical results confirm that the GH shift provides a direct real-space probe for the bulk Floquet topology.

\section{Conclusion and Discussion}\label{secvi}

In summary, we have investigated the boundary scattering dynamics of two-dimensional Floquet topological insulators. Using a discrete-time scattering formalism, we established an exact equivalence between the winding number of the reflection phase $\nu(\epsilon)$ and the bulk Floquet invariant $W_3(\epsilon)$ for a given quasienergy gap $\epsilon$. Under transverse open boundary conditions, the reflection matrix exhibits the NHSE: the eigenspectrum collapses inside the unit circle, and the corresponding skin modes localize at the interface corners. This behavior arises from the unidirectional wave transport associated with the chiral edge states. Furthermore, we demonstrated the gap-dependence of the GH shift for an incident wave packet and showed that it provides a direct real-space measure of the topological invariant for the targeted gap.

Our results offer a practical boundary-based approach to probe the dynamical topology of the bulk. Unlike momentum-space tomography of the bulk quasienergy bands, which may be challenging for complex driving protocols, measuring the GH shift relies solely on boundary scattering. Such $k_y$-resolved measurements are highly feasible in platforms like acoustic metamaterials, photonic platforms, and ultracold atoms, where wave packet preparation and spatial shift detection are readily accessible \cite{ Aidelsburger2015, Zeuner2015, Wimmer2017}. Experimentally, each quasienergy gap can be independently probed by tuning the incident energy $\epsilon_0$ of the wave packet. In cold atoms, for example, by varying the incident angle---which corresponds to tuning the initial transverse momentum $k_{y0}$ of the wave packet---and recording the resulting transverse shift $\Delta_{GH}(\epsilon_0, k_{y0})$, the topological invariant can be unambiguously extracted. As the scattering is formulated in real space, our method can be applied to disordered Floquet systems where the bulk momentum ceases to be a good quantum number. Given the topological robustness of the chiral edge modes and the skin effect \cite{Hu_Lyapunov}, we expect the GH shift to persist under weak disorder. The fate of the point-gap topology and the reflection skin effect in the presence of strong disorder warrants further investigation. Additionally, generalizing this framework to interacting driven systems remains an open question.

\begin{acknowledgements}
 This work is supported by the National Key Research and Development Program of China (Grants No. 2022YFA1405800 and No. 2023YFA1406704) and National Natural Science Foundation of China (Grant No. 12474496 and No. 12547107).
\end{acknowledgements}


\begin{thebibliography}{99}

\bibitem{Bukov2015} M. Bukov, L. D'Alessio, and A. Polkovnikov, Universal high-frequency behavior of periodically driven systems: from dynamical stabilization to Floquet engineering, \href{https://doi.org/10.1080/00018732.2015.1055918}{Adv. Phys. {\bf 64}, 139 (2015)}.
\bibitem{Eckardt2017} A. Eckardt, Colloquium: Atomic quantum gases in periodically driven optical lattices, \href{https://doi.org/10.1103/RevModPhys.89.011004}{Rev. Mod. Phys. {\bf 89}, 011004 (2017)}.
\bibitem{Oka2019} T. Oka and S. Kitagawa, Floquet Engineering of Quantum Materials, \href{https://doi.org/10.1146/annurev-conmatphys-031218-013423}{Annu. Rev. Condens. Matter Phys. {\bf 10}, 387 (2019)}.
\bibitem{Oka2009} T. Oka and H. Aoki, Photovoltaic Hall effect in graphene, \href{https://doi.org/10.1103/PhysRevB.79.081406}{Phys. Rev. B {\bf 79}, 081406 (2009)}.
\bibitem{Kitagawa2010} T. Kitagawa, E. Berg, M. Rudner, and E. Demler, Topological phases of periodically driven systems, \href{https://doi.org/10.1103/PhysRevB.82.235114}{Phys. Rev. B {\bf 82}, 235114 (2010)}.
\bibitem{Lindner2011} N. H. Lindner, G. Refael, and V. Galitski, Floquet topological insulator in semiconductor quantum wells, \href{https://doi.org/10.1038/nphys1926}{Nat. Phys. {\bf 7}, 490 (2011)}.
\bibitem{Jiang2011} L. Jiang, T. Kitagawa, J. Alicea, A. R. Akhmerov, D. Pekker, G. Refael, J. I. Cirac, E. Demler, M. D. Lukin, and P. Zoller, Majorana Fermions in Equilibrium and in Driven Cold-Atom Quantum Wires, \href{https://doi.org/10.1103/PhysRevLett.106.220402}{Phys. Rev. Lett. {\bf 106}, 220402 (2011)}.
\bibitem{Cayssol2013} J. Cayssol, B. Dora, F. Simon, and R. Moessner, Floquet Topological Insulators, \href{https://doi.org/10.1002/pssr.201206451}{Phys. Status Solidi RRL {\bf 7}, 101 (2013)}.
\bibitem{Goldman2014} N. Goldman and J. Dalibard, Periodically Driven Quantum Systems: Effective Hamiltonians and Engineered Gauge Fields, \href{https://doi.org/10.1103/PhysRevX.4.031027}{Phys. Rev. X {\bf 4}, 031027 (2014)}.
\bibitem{Usaj2014} G. Usaj, P. M. Perez-Piskunow, L. E. F. Foa Torres, and C. A. Balseiro, Irradiated graphene as a tunable Floquet topological insulator, \href{https://doi.org/10.1103/PhysRevB.90.115423}{Phys. Rev. B {\bf 90}, 115423 (2014)}.

\bibitem{Wang2013} Y. H. Wang, H. Steinberg, P. Jarillo-Herrero, and N. Gedik, Observation of Floquet-Bloch States on the Surface of a Topological Insulator, \href{https://doi.org/10.1126/science.1239834}{Science {\bf 342}, 453 (2013)}.
\bibitem{McIver2020} J. W. McIver, B. Schulte, F.-U. Stein, T. Matsuyama, G. Jotzu, G. Meier, and A. Cavalleri, Light-induced anomalous Hall effect in graphene, \href{https://doi.org/10.1038/s41567-019-0698-y}{Nat. Phys. {\bf 16}, 38 (2020)}.
\bibitem{Roushan2017} P. Roushan et al., Chiral ground-state currents of interacting photons in a synthetic magnetic field, \href{https://doi.org/10.1038/nphys3930}{Nat. Phys. {\bf 13}, 146 (2017)}.
\bibitem{Peng2016} Y.-G. Peng, C.-Z. Qin, D.-G. Zhao, Y.-X. Shen, X.-Y. Xu, M. Bao, H. Jia, and X.-F. Zhu, Experimental demonstration of anomalous Floquet topological insulator for sound, \href{https://doi.org/10.1038/ncomms13368}{Nat. Commun. {\bf 7}, 13368 (2016)}.

\bibitem{Jotzu2014} G. Jotzu et al., Experimental realization of the topological Haldane model with ultracold fermions, \href{https://doi.org/10.1038/nature13915}{Nature {\bf 515}, 237 (2014)}.
\bibitem{Aidelsburger2015} M. Aidelsburger, M. Lohse, C. Schweizer, M. Atala, J. T. Barreiro, S. Nascimbène, N. R. Cooper, I. Bloch, and N. Goldman, Measuring the Chern number of Hofstadter bands with ultracold bosonic atoms, \href{https://doi.org/10.1038/nphys3171}{Nat. Phys. {\bf 11}, 162 (2015)}.
\bibitem{Wintersperger2020} K. Wintersperger et al., Realization of an anomalous Floquet topological system with ultracold atoms, \href{https://doi.org/10.1038/s41567-020-0949-y}{Nat. Phys. {\bf 16}, 1058 (2020)}.

\bibitem{Rechtsman2013} M. C. Rechtsman et al., Photonic Floquet topological insulators, \href{https://doi.org/10.1038/nature12066}{Nature {\bf 496}, 196 (2013)}.
\bibitem{Maczewsky2017} L. J. Maczewsky et al., Observation of photonic anomalous Floquet topological insulators, \href{https://doi.org/10.1038/ncomms13756}{Nat. Commun. {\bf 8}, 13756 (2017)}.
\bibitem{Mukherjee2017} S. Mukherjee, A. Spracklen, M. Valiente, E. Andersson, P. Öhberg, N. Goldman, and R. R. Thomson, Experimental observation of anomalous topological edge modes in a slowly driven photonic lattice, \href{https://doi.org/10.1038/ncomms13918}{Nat. Commun. {\bf 8}, 13918 (2017)}.

\bibitem{Rudner2013} M. S. Rudner, N. H. Lindner, G. Refael, and V. Galitski, Anomalous Edge States and the Bulk-Edge Correspondence for Periodically Driven 2D Systems, \href{https://doi.org/10.1103/PhysRevX.3.031005}{Phys. Rev. X {\bf 3}, 031005 (2013)}.
\bibitem{Titum2016} P. Titum, E. Berg, M. S. Rudner, G. Refael, and N. H. Lindner, Anomalous Floquet-Anderson Insulator as a Nonadiabatic Quantized Charge Pump, \href{https://doi.org/10.1103/PhysRevX.6.021013}{Phys. Rev. X {\bf 6}, 021013 (2016)}.
\bibitem{Roy2017} R. Roy and F. Harper, Periodic table for Floquet topological insulators, \href{https://doi.org/10.1103/PhysRevB.96.155118}{Phys. Rev. B {\bf 96}, 155118 (2017)}.
\bibitem{Yao2017} S. Yao, Z. Yan, and Z. Wang, Topological invariants of Floquet systems: General formulation, special properties, and Floquet topological defects, \href{https://doi.org/10.1103/PhysRevB.96.195303}{Phys. Rev. B {\bf 96}, 195303 (2017)}.
\bibitem{Nathan2015} F. Nathan and M. S. Rudner, Topological singularities and the general classification of Floquet-Bloch systems, \href{https://doi.org/10.1088/1367-2630/17/12/125014}{New J. Phys. {\bf 17}, 125014 (2015)}.

\bibitem{Carpentier2015} D. Carpentier, P. Delplace, M. Fruchart, and K. Gawedzki, Topological Index for Periodically Driven Time-Reversal Invariant 2D Systems, \href{https://doi.org/10.1103/PhysRevLett.114.106806}{Phys. Rev. Lett. {\bf 114}, 106806 (2015)}.
\bibitem{Morimoto2017} T. Morimoto, H. C. Po, and A. Vishwanath, Floquet topological phases protected by time glide symmetry, \href{https://doi.org/10.1103/PhysRevB.95.195155}{Phys. Rev. B {\bf 95}, 195155 (2017)}.
\bibitem{Hu_2020} H. Hu, B. Huang, E. Zhao, and W. V. Liu, Dynamical Singularities of Floquet Higher-Order Topological Insulators, \href{https://doi.org/10.1103/PhysRevLett.124.057001}{Phys. Rev. Lett. {\bf 124}, 057001 (2020).}
\bibitem{Huang_2020} B. Huang and W. V. Liu, Floquet Higher-Order Topological Insulators with Anomalous Dynamical Polarization, \href{https://doi.org/10.1103/PhysRevLett.124.216601}{Physical Review Letters {\bf 124}, 216601 (2020).}

\bibitem{Asboth2014} J. K. Asbóth, B. Tarasinski, and P. Delplace, Chiral symmetry and bulk-boundary correspondence in periodically driven one-dimensional systems, \href{https://doi.org/10.1103/PhysRevB.90.125143}{Phys. Rev. B {\bf 90}, 125143 (2014)}.
\bibitem{Fruchart2016} M. Fruchart, Complex classes of Floquet topological insulators, \href{https://doi.org/10.1103/PhysRevB.93.115429}{Phys. Rev. B {\bf 93}, 115429 (2016)}.

\bibitem{Hauke2014} P. Hauke, M. Lewenstein, and A. Eckardt, Tomography of band insulators from quench dynamics, \href{https://doi.org/10.1103/PhysRevLett.113.045303}{Phys. Rev. Lett. {\bf 113}, 045303 (2014)}.
\bibitem{Flaschner2016} N. Fläschner et al., Experimental reconstruction of the Berry curvature in a Floquet Bloch band, \href{https://doi.org/10.1126/science.aad4568}{Science {\bf 352}, 1091 (2016)}.
\bibitem{Unal2019} F. N. Ünal, B. Seradjeh, and A. Eckardt, How to Directly Measure Floquet Topological Invariants in Optical Lattices, \href{https://doi.org/10.1103/PhysRevLett.122.253601}{Phys. Rev. Lett. {\bf 122}, 253601 (2019)}.
\bibitem{Asteria2019} L. Asteria, D. T. Tran, T. Ozawa, M. Tarnowski, B. S. Rem, N. Fläschner, K. Sengstock, N. Goldman, and C. Weitenberg, Measuring quantized circular dichroism in ultracold topological matter, \href{https://doi.org/10.1038/s41567-019-0417-8}{Nat. Phys. {\bf 15}, 449 (2019).}

\bibitem{Fulga2012} I. C. Fulga, F. Hassler, and A. R. Akhmerov, Scattering theory of topological insulators and superconductors, \href{https://doi.org/10.1103/PhysRevB.85.165409}{Phys. Rev. B {\bf 85}, 165409 (2012)}.
\bibitem{Franca2022} S. Franca, V. Könye, F. Hassler, J. van den Brink, and C. Fulga, Non-Hermitian physics without gain or loss: the skin effect of reflected waves, \href{https://doi.org/10.1103/PhysRevLett.129.086601}{Phys. Rev. Lett. {\bf 129}, 086601 (2022)}.

\bibitem{Yao2018} S. Yao and Z. Wang, Edge States and Topological Invariants of Non-Hermitian Systems, \href{https://doi.org/10.1103/PhysRevLett.121.086803}{Phys. Rev. Lett. {\bf 121}, 086803 (2018)}.
\bibitem{MartinezAlvarez2018} V. M. Martinez Alvarez, J. E. Barrios Vargas, and L. E. F. Foa Torres, Non-Hermitian robust edge states in one dimension: Anomalous localization and eigenspace condensation at exceptional points, \href{https://doi.org/10.1103/PhysRevB.97.121401}{Phys. Rev. B {\bf 97}, 121401(R) (2018)}.
\bibitem{Yokomizo2019} K. Yokomizo and S. Murakami, Non-Bloch Band Theory of Non-Hermitian Systems, \href{https://doi.org/10.1103/PhysRevLett.123.066404}{Phys. Rev. Lett. {\bf 123}, 066404 (2019)}.
\bibitem{Lee2019} C. H. Lee and R. Thomale, Anatomy of skin modes and topology in non-Hermitian systems, \href{https://doi.org/10.1103/PhysRevB.99.201103}{Phys. Rev. B {\bf 99}, 201103(R) (2019)}.
\bibitem{Zhang2020} K. Zhang, Z. Yang, and C. Fang, Correspondence between Winding Numbers and Skin Modes in Non-Hermitian Systems, \href{https://doi.org/10.1103/PhysRevLett.125.126402}{Phys. Rev. Lett. {\bf 125}, 126402 (2020)}.
\bibitem{Okuma2020} N. Okuma, K. Kawabata, K. Shiozaki, and M. Sato, Topological Origin of Non-Hermitian Skin Effects, \href{https://doi.org/10.1103/PhysRevLett.124.086801}{Phys. Rev. Lett. {\bf 124}, 086801 (2020)}.
\bibitem{Kunst2018} F. K. Kunst, E. Edvardsson, J. C. Budich, and E. J. Bergholtz, Biorthogonal Bulk-Boundary Correspondence in Non-Hermitian Systems, \href{https://doi.org/10.1103/PhysRevLett.121.026808}{Phys. Rev. Lett. {\bf 121}, 026808 (2018)}.
\bibitem{Borgnia2020} D. S. Borgnia, A. J. Kruchkov, and R.-J. Slager, Non-Hermitian Boundary Modes and Topology, \href{https://doi.org/10.1103/PhysRevLett.124.056802}{Phys. Rev. Lett. {\bf 124}, 056802 (2020)}.
\bibitem{Hu_2025} H. Hu, Topological origin of non-Hermitian skin effect in higher dimensions and uniform spectra, \href{https://doi.org/10.1016/j.scib.2024.07.022}{Science Bulletin {\bf 70}, 51 (2025).}

\bibitem{Ma2020} H. Ma, C. Ju, X. Xi, and R.-X. Wu, Nonreciprocal Goos-Hänchen shift by topological edge states of a magnetic photonic crystal, \href{https://doi.org/10.1364/OE.394286}{Opt. Express {\bf 28}, 19916 (2020)}.

\bibitem{Fulga2016} I. C. Fulga and M. Maksymenko, Scattering matrix invariants of Floquet topological insulators, \href{https://doi.org/10.1103/PhysRevB.93.075405}{Phys. Rev. B {\bf 93}, 075405 (2016)}.
\bibitem{Chalker1988} J. T. Chalker and P. D. Coddington, Percolation, quantum tunnelling and the integer Hall effect, \href{https://doi.org/10.1088/0022-3719/21/14/008}{J. Phys. C: Solid State Phys. {\bf 21}, 2665 (1988)}.
\bibitem{Tajic2005} A. Tajic, Study of a stroboscopic model of a quantum dot, Ph.D. thesis, Leiden University (2005).

\bibitem{Gong2018} Z. Gong, Y. Ashida, K. Kawabata, K. Takasan, S. Higashikawa, and M. Ueda, Topological Phases of Non-Hermitian Systems, \href{https://doi.org/10.1103/PhysRevX.8.031079}{Phys. Rev. X {\bf 8}, 031079 (2018)}.
\bibitem{Kawabata2019} K. Kawabata, K. Shiozaki, M. Ueda, and M. Sato, Symmetry and Topology in Non-Hermitian Physics, \href{https://doi.org/10.1103/PhysRevX.9.041015}{Phys. Rev. X {\bf 9}, 041015 (2019)}.

\bibitem{Beenakker2009} C. W. J. Beenakker, R. A. Sepkhanov, A. R. Akhmerov, and J. Tworzyd{\l}o, Quantum Goos-H\"anchen Effect in Graphene, \href{https://doi.org/10.1103/PhysRevLett.102.146804}{Phys. Rev. Lett. {\bf 102}, 146804 (2009)}.

\bibitem{Bliokh2013} K. Y. Bliokh and A. Aiello, Goos-H\"anchen and Imbert-Fedorov beam shifts: An overview, \href{https://doi.org/10.1088/2040-8978/15/1/014001}{J. Opt. {\bf 15}, 014001 (2013)}.

\bibitem{Zeuner2015} J. M. Zeuner, M. C. Rechtsman, Y. Plotnik, Y. Lumer, S. Nolte, M. S. Segev, and A. Szameit, Observation of a Topological Transition in the Bulk of a Non-Hermitian System, \href{https://doi.org/10.1103/PhysRevLett.115.040402}{Phys. Rev. Lett. {\bf 115}, 040402 (2015)}.

\bibitem{Wimmer2017} M. Wimmer, H. M. Price, I. Carusotto, and U. Peschel, Experimental measurement of the Berry curvature from anomalous transport, \href{https://doi.org/10.1038/nphys4050}{Nat. Phys. {\bf 13}, 545 (2017)}.

\bibitem{Hu_Lyapunov} K. Sun and H. Hu, Lyapunov formulation of band theory for disordered non-Hermitian systems, \href{https://doi.org/10.48550/arXiv.2507.09447}{arXiv.2507.09447.}
\end{thebibliography}
\end{document}